# Atomic-scale grain boundary engineering to overcome hot-cracking in additively-manufactured superalloys

Paraskevas Kontis[1*], Edouard Chauvet[2,3], Zirong Peng[1], Junyang He[1], Alisson Kwiatkowski da Silva[1], Dierk Raabe[1], Catherine Tassin[2], Jean-Jacques Blandin[2], Stéphane Abed[3], Rémy Dendievel[2], Baptiste Gault[1,4], Guilhem Martin[2*]

[1] Max-Planck-Institut für Eisenforschung, Max-Planck-Str. 1, 40237 Düsseldorf, Germany.

[2] Univ. Grenoble Alpes, CNRS, Grenoble INP, SIMaP, F-38000 Grenoble, France.

[3] Poly-Shape, 235 Rue des Canesteu -ZI La Gandonne, 13300, Salon-de-Provence, France.

[4] Department of Materials, Imperial College London, South Kensington, SW7 2AZ, UK.

## Abstract

There are still debates regarding the mechanisms that lead to hot cracking in parts build by additive manufacturing (AM) of non-weldable nickel-based superalloys. This lack of in-depth understanding of the root causes of hot cracking is an impediment to designing engineering parts for safety-critical applications. Here, we deploy a near-atomic-scale approach to investigate the details of the compositional decoration of grain boundaries in the coarse-grained, columnar microstructure in parts built from a non-weldable nickel-based superalloy by selective electron-beam melting. The progressive enrichment in Cr, Mo and B at grain boundaries over the course of the AM-typical successive solidification and remelting events, accompanied by solid-state diffusion, causes grain boundary segregation induced liquation. This observation is consistent with thermodynamic calculations. We demonstrate that by adjusting build parameters to obtain a fine-grained equiaxed or a columnar microstructure with grain width smaller than 100 μm enables to avoid cracking, despite strong grain boundary segregation. We find that the spread of critical solutes to a higher total interfacial area, combined with lower thermal stresses, helps to suppress interfacial liquation.

**Keywords:** Hot cracking, grain boundaries, electron beam melting, superalloys, liquid film

[*] Corresponding authors: **p.kontis@mpie.de, guilhem.martin@simap.grenoble-inp.fr**

## 1 Introduction

Additive manufacturing (AM) has a great potential for the production of novel engineering components for aerospace and power generation applications, where nickel-based superalloys are predominantly used [1]. However, it is very challenging to produce safety-critical components by AM with zero-crack tolerance, yet the digitalisation inherent to AM promises large gains in manufacturing and repair efficiency [2]. The AM community currently invests great efforts to 3D print well-known superalloys compositions considered as non-weldable such as IN738 [3–5], CMSX-4 [6], CM247LC [7] among others [8,9]. These alloy grades, which have been classified as non-weldable due to their Ti and Al contents [2] usually imply formation of large volume fractions of γ'-precipitates and were observed to undergo severe cracking when welded or AM built.

Amongst the possible technologies to fabricate parts highly susceptible to cracking, selective electron-beam melting (S-EBM) is one of the most promising. The main advantage is the pre-heated powder bed (typically around 1000°C for nickel-based superalloys) resulting in the reduction of thermal stresses arising from cooling. This pre-heating however makes the thermal history of the sample more complex and sometimes difficult to control. In addition, S-EBM offers precise control over the melting strategy and hence enables tuning of the microstructure by modifying the thermal gradient and solidification velocity. Pragmatic approaches have been deployed, mainly on the weldable Inconel 718 grade, to adjust parameters to modify the direction of the main thermal gradient with respect to the building direction and the solidification velocity [10–17]. Successful attempts have been reported in the literature. Helmer et al. [10,11] showed that the grain size and grain morphology can be modified by appropriate changes of the melting parameters. Dehoff et al. [12,13] have demonstrated that a spatially controlled crystallographic texture within Inconel 718 components can be achieved. Such microstructure engineering approach was further investigated by Koepf et al. [14,15] and Rhagavan et al. [16,17].

Cracks developing during S-EBM of non-weldable alloys are often qualified as hot cracks because cracking occurs in presence of liquid films, as is the case for the grade investigated in the present work [18]. The combination of liquid films and thermal stresses arising from shrinkage and cooling leads to the development of hot cracks. Hot cracks were found to systematically propagate along high angle grain boundaries (HAGBs) while low angle grain boundaries (LAGBs) were not susceptible to hot cracking, similarly to what was reported in the welding literature [19]. This difference of cracking sensitivity between HAGBs and LAGBs was rationalized using the theoretical model of Rappaz et al. [20] that describes the last stage of solidification. Relying on the concept of repulsive (stabilization of liquid films) and attractive boundaries (dendrite coalescence), it was explained that the coalescence undercooling was larger for HAGBs than for LAGBs, which leads to liquid films stable at lower temperatures along HAGBs. The latter observation suggests that producing samples without HAGBs can be considered as a possible route to achieve crack-free parts. Körner et al. [21] and Chauvet et al. [22] took advantage of the digital control of the melting strategy to grow single crystals of non-weldable nickel-based superalloys by S-EBM. Interestingly, the single crystals grown by S-EBM are free of cracks, in agreement with the observation that cracks propagate exclusively along HAGBs. However, in the work presented in ref. [18], the exact mechanism leading to cracking observed along HAGBs was still under debate, i.e. solidification cracking *vs.* liquation cracking. Identifying the root causes of cracking requires a detailed investigation of solutes segregating along HAGBs at the near-atomic scale. Understanding of the underlying interface decoration and decohesion mechanisms should thus help to guide strategies to avoid the development of those critical defects and allow to produce crack-free parts.

Here, we perform a detailed microstructural characterization by scanning electron microscopy (SEM) and atom probe tomography (APT) of cuboidal samples fabricated by S-EBM from non-weldable crack sensitive superalloy powders [8], which we use to guide thermodynamic calculations. Our new insights enable us to identify the atomic-scale cause of cracking along HAGBs in the columnar microstructure formed by S-EBM. We show that cracks are caused by segregation-induced liquation at HAGBs decorated by segregated films enriched in Cr, Mo and B. Interestingly, B is originally added to enhance grain boundary cohesion. The flexibility in processing offered by S-EBM allows us to demonstrate that by controlling the interface area, in particular the grain boundary density, we can limit solute enrichment in the intergranular region and prevent critical liquation, alleviating cracking susceptibility. Increasing drastically the interface area not only helps to limit solute enrichment at HAGBs but also to contribute to better distribute thermal stresses that act as the mechanical driving force for hot cracking. This grain boundary engineering strategy is demonstrated in two new crack-free builds, respectively exhibiting an equiaxed and a fine-columnar microstructure (grain width < 100 μm). Shedding light on fundamental aspects of cracking allows us to derive atomic-scale inspired digital metallurgy design pathways to enable application of these alloys in safety-critical applications, such as blades in aero-engines.

## 2 Experimental procedures

### 2.1 Powder characteristics

The material investigated is a non-weldable nickel-based superalloy. The pre-alloyed powder was produced by gas atomization and provided by ERASTEEL. The powder composition was measured by thermal-conductivity-infrared for the C and O, and by inductively-coupled plasma- atomic emission spectroscopy (ICP-AES) for the other elements. The chemical composition is reported in at.% in **Table 1**, listing the most relevant key elements.

It is known that in some cases the bulk composition can be altered after the deposition, such as in the case of Al which is sensitive to evaporation when processing Ti-6Al-4V alloy by S-EBM [23]. However, chemical analysis performed on the produced bulk samples revealed no substantial alteration of the bulk composition after S-EBM processing. The as-received powder batch exhibits mostly a spherical morphology with some satellites. The powder characteristics are summarized as follows: $D_{50}$ = 75 μm, a powder bed density of 53.6% and a flow time of 16.0 ± 0.1 s for 50 g of powders going through a 2.54 mm diameter orifice (Hall flowmeter).

Table 1: Chemical composition of the powders used in the present investigation in at.%.

| Ni | Cr | Mo | Co | Ti | Al | Zr | Hf | B | C | Si |
|---|---|---|---|---|---|---|---|---|---|---|
| Bal. | 12.20 | 3.80 | 14.85 | 5.30 | 8.70 | 0.02-0.025 | 0.18-0.21 | 0.083 | 0.06-0.07 | 0.04 |

### 2.2 Selective electron beam melting

Samples with dimensions 23 x 23 x 30 mm (width x length x height) were built in an ARCAM A1 Electron Beam Melting machine operating at 60 kV accelerating voltage under a controlled pressure of He set to $2.10^{-3}$ mbar, with the layer thickness set to 50 μm. Two different versions of the same composition were built. The first one was produced using the automatic mode while the second one was melted using the manual mode. The automatic mode resulted in a sample with relatively coarse columnar-shaped grains and it will be denoted from now as cracked columnar sample. The second sample was built with a higher scanning speed and a reduced hatch spacing resulting in a reduced linear energy (0.119 J/m²) when compared to the one employed in the automated mode (0.5-0.65 J/mm²). The significant overlap of the beam scanned multiple times over adjacent locations leads to an elongated melt pool perpendicularly to the hatching [10,11]. This hinders epitaxial growth of new grains on the grains from the layer underneath and leads to an equiaxed microstructure [11,24]. This sample is denoted as crack-free with equiaxed grains. All the used processing conditions were preliminary optimized to achieve dense samples (density >99.8%) with no lack-of-fusion defects. **Table 2** summarizes the main parameters used to build the two samples. The building direction is denoted as Z.

A build plate made of a polycrystalline hot-rolled 10 mm thick stainless steel plate, grade EN 1.4307 was used during the production of all the samples. Although, its mechanical resistance i.e. creep, is sufficient to allow us to run stable builds, after depositing the samples it was slightly deformed. Diffusion between the substrate and the deposited material likely occurs over limited distance, however in this study only the upper parts of the samples were investigated which are not affected by this diffusion.

In each case, the build plate was heated at 1050°C and the preheating temperature was maintained at 1050°C throughout the build and the focus offset set to 20 mA during the melting steps. When operated in automatic mode, the melting parameters, such as caused by the beam power and the scanning speed, were adjusted as a function of the geometry through the use of automated functions such as the *speed function* or *turning point function* into the software EBM Control 3.2, see **Table 2** for more details. In the case of the manual mode, the power and scan speed were imposed and do not depend on the geometry. In both cases, the melting strategy consisted in scanning the area defined by the CAD in a snake-like manner with a line order set to 1 and the scanning direction was also rotated by 90° after every layer.

Table 2: Summary of the main processing parameters used for building the three samples.

|  | **Cracked columnar** | **Crack-free with equiaxed grains** |
|---|---|---|
| **Mode** | Automatic* | Manual |
| **Cracks** | Yes (upper part Z >12-15mm) | No |
| **Preheating Temperature, $T_p$ (°C)** | 1050 | 1050 |
| **Linear Energy = $P/v$ (J/mm²)** | Auto* (0.5-0.65) | 0.119 |
| **Line Order** | 1 | 1 |
| **Focus Offset, *FO* (mA)** | 20 | 20 |

*Power and scanning speed automatically determined by the thermal model used in the EBM Control 3.2 software which depends on the geometry of the parts to be built. When running the automatic mode of the EBM control 3.2 software, the scanning speed changes spatially (x, y, z) through the use of automated functions:
 - Speed function: a function that changes automatically the scan speed as a function of the beam power. The higher the speed function, the higher the scan speed. For cuboidal samples, the scan length is constant, hence the beam power is constant.
 - Turning point function: a function that further changes the beam speed to account for edge effects. To reduce the applied energy near the edge, this function is used to increase the beam speed as it comes to an edge.
 - Thickness function: a function that modifies the beam speed to compensate for the lower amount of energy required in overhangs based on the distance down to powder. This function was not operating for cuboidal geometries.

## 2.3 Mechanical testing

Micro-specimens with a gauge length of 3 mm, a width of 1 mm, and a thickness of 0.8 mm were machined by Electron Discharge Machining (EDM) from the as-built S-EBM samples, with the load axis normal to the building direction Z. Tensile tests were performed at room temperature using an ADAMEL DY35 universal testing machine equipped with a 2kN load cell at a strain rate of $5.10^{-4}$ $s^{-1}$. The strain was measured by digital image correlation (GOM-ARAMIS software). One side of the gauge length of the micro-tensile samples was preliminary covered with a speckle made of black and white paintings. Images were recorded during the tensile tests at a frequency of 1 Hz.

## 2.4 Scanning electron microscopy

The as-build samples were cut along the build direction Z and they were mechanically polished with abrasive media to a 0.04 μm colloidal silica finish for microstructural characterization. Optical microscopy was used for observations at large scale whereas a Zeiss Merlin field emission gun scanning electron microscope (FEG-SEM) was used in backscattered mode for fine microstructural observations. Electron backscattered (EBSD) analysis was performed in a Zeiss Ultra FEG-SEM operated at 20kV acceleration voltage with a step size of 2 μm. Data for maps of 1mm$^2$ size were collected using the TSL-OIM software. A grain was defined as a group of pixels containing at least 10 pixels, including multiple rows, and having a tolerance angle of 5°.

## 2.5 Atom probe tomography

In order to study the segregation of solutes at grain boundaries, specimens for atom probe tomography (APT) were prepared by following a site specific lift-out protocol outlined in Ref. [25], using a FEI Dual Beam FIB Helios 600i. The APT specimens were extracted from random high angle grain boundaries from both samples, following the procedures described in Ref. [26]. APT specimens were analyzed using a Cameca LEAP 5000 XR instrument operating in a laser mode with pulse rate at 125 kHz, pulse energy 0.4 pJ and temperature 60 K. The commercial package IVAS 3.8.2 was used for data reconstruction and analyses. In addition, the segregation at grain boundaries as observed by APT was further analyzed by making use of the method described in Ref. [27].

Finally, in the case of the sample with equiaxed grains, correlative Transmission Kikuchi diffraction (TKD) and APT analysis was performed on the APT specimens to confirm the presence of a grain boundary prior to APT experiments. TKD data were obtained with a Zeiss Merlin FEG-SEM equipped with a Bruker Optimus detector. For the measurement acceleration voltage of 30kV and probe current of 1.5nA was used with a step size of 5nm at 70 fps.

## 2.6 Thermodynamic calculations

To evaluate the effect of the content of B, Cr and Mo on the solidification path, thermodynamic calculations were performed using the Thermo-Calc software making use of the TTNI8 database dedicated to nickel-based superalloys [28]. Calculations of the solidification path assuming constitutive segregation based on the Scheil-Gulliver model were performed including the liquid, γ and γ' phases considering both, excluding and including the formation of $M_2B$ borides. No diffusion in the solid state and infinite diffusion in the liquid was assumed for the Scheil-Gulliver model.

## 3 Results

### 3.1 Microstructural characterization: cracked columnar sample

The first build, achieved by operating in automatic mode, is shown in **Figure 1(a)**. A columnar microstructure develops with numerous thin grains, which progressively becoming coarser from approximately the building height of 5 mm, as evidenced by the EBSD maps in **Figure 1(b-d)**. The sample exhibits a strong <001>-fiber texture typical of directionally solidified superalloys and FCC alloys. Cracked HAGBs are often observed (**Figure 1(e)**)in the upper part of the build (height > 10 mm), (**Figure 1(f)**). The opened fracture surface in **Figure 1(f)** reveals a dendritic morphology typical of hot cracking [20], which implies that a continuous liquid film had covered almost the sample's entire cross section.

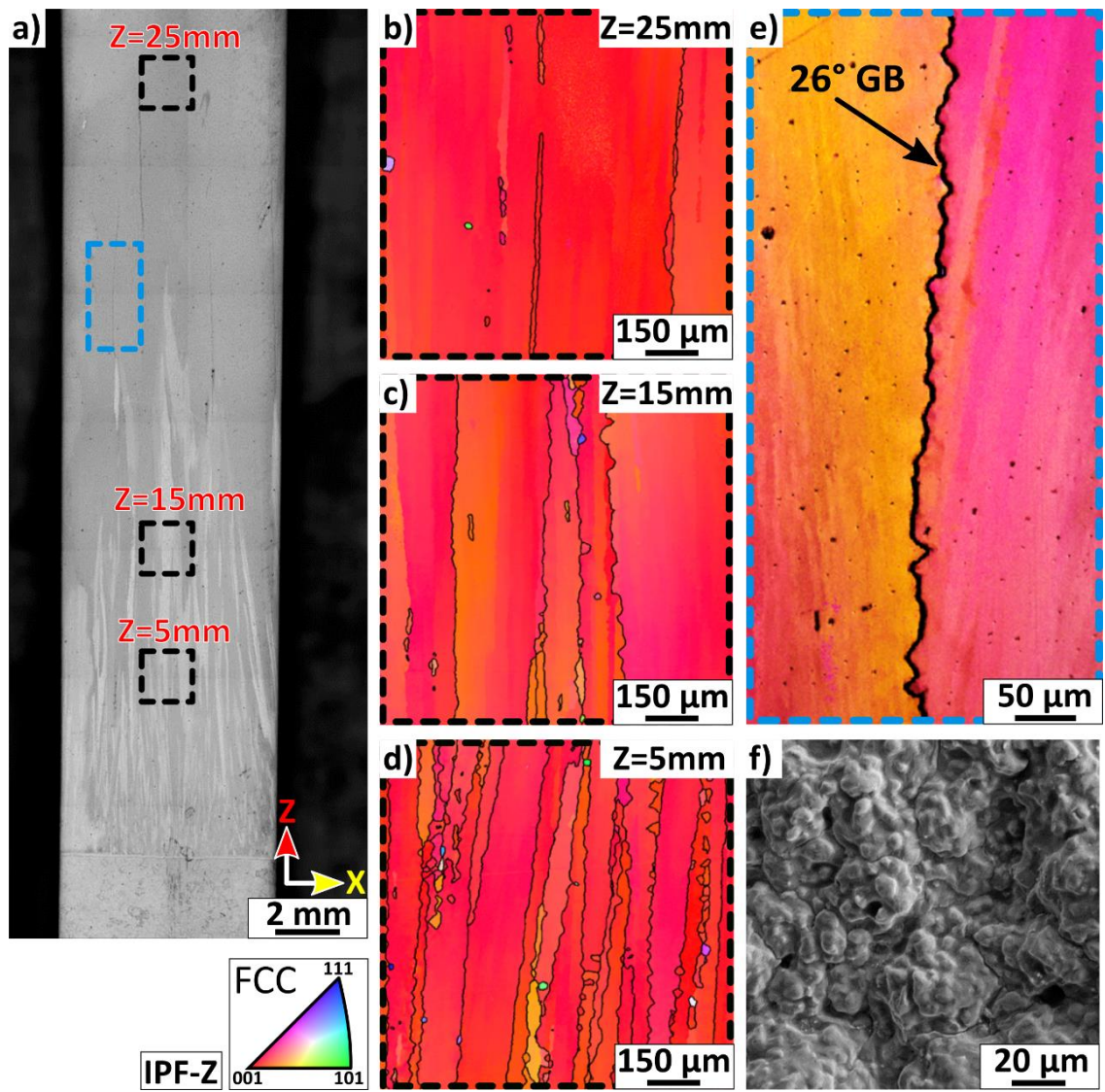

Figure 1: Overview of the cracked columnar sample. a) Optical micrograph of the entire height showing a graded microstructure along the building direction. b)-d) EBSD IPF-Z (building direction) maps illustrating the columnar grain width evolution along the building direction and the strong crystallographic fiber texture. e) Enlarged view of a HAGB (misorientation 26°) affected by hot cracking in the upper region of the specimen. f) Fracture surface after tensile testing giving evidence of the presence of liquid films.

Cracking only takes place at high-angle grain boundaries [21] and building height greater than 10 mm in the build, where the columnar grains are wider than 100 μm. LAGBs, consisting of dislocation arrays, are systematically found to be crack-free. Importantly, cracks do not form or propagate through the final melted layer. This means that the final melted layer is the only region where the microstructure inherited from solidification is unaffected by subsequent layers or preheating at the top of the build, as revealed by **Figure 2**.

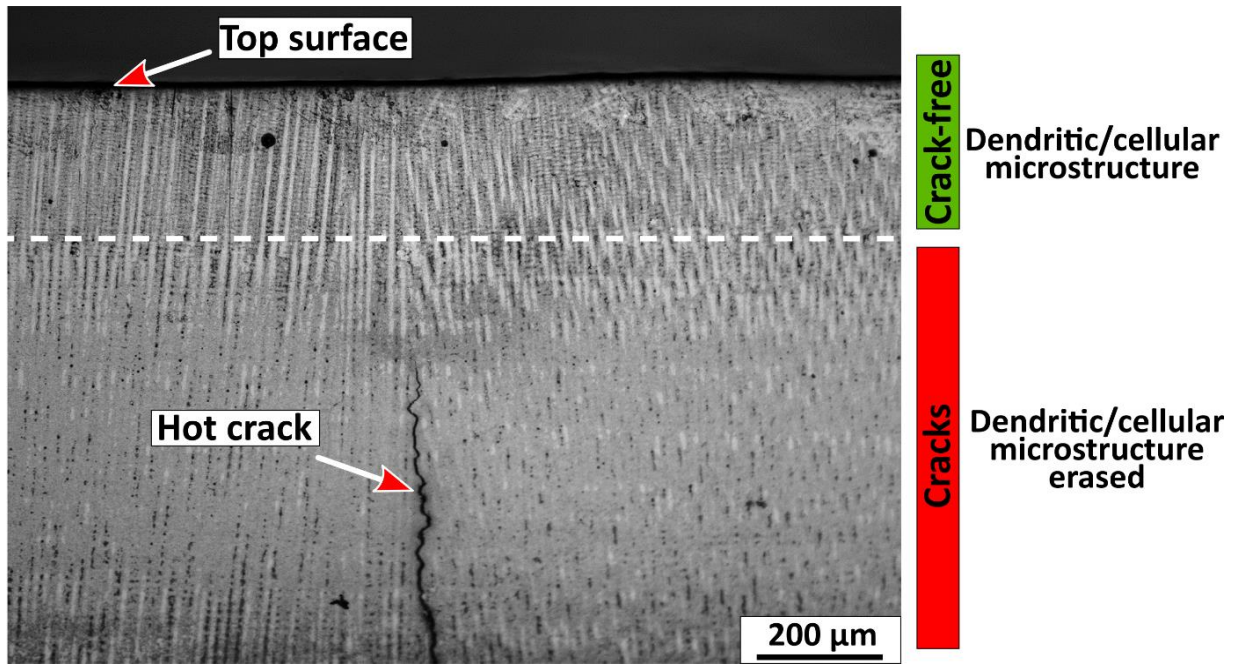

Figure 2: Optical micrograph towards the top of the cracked columnar sample highlighting that cracks do not propagate through the last melted layer. Note that the dendritic microstructure can only be seen in the final melted region while it is erased in the underlying layer because the complex thermal history acts as a homogenization treatment.

Particles with bright contrast are often observed at cracked HAGBs in backscattered electron micrographs. **Figure 3(a-c)** shows a decorated HAGB between grains containing the typical γ/γ' microstructure. Atom probe tomography (APT) reveals that these particles are $M_2B$ borides, containing almost 40 at.% Mo and 20 at.% Cr as reported in **Table 3**, consistent with previous reports [18,29]. Not all HAGBs appear cracked, but most HAGBs bearing coarse borides are cracked. By contrast, LAGBs were not found to be cracked and were not decorated by $M_2B$-borides. **Figure 3(a)** shows for instance cracks on either side of the region along the boundary where the boride formed and coarsened. **Figure 3(d)** shows that recrystallized regions exist near the borides, which are the sign of either high thermal stresses causing local plastic deformation, thereby storing dislocations that subsequently induce recrystallization, or of the coarsening of the borides [30,31]. The local grain boundary composition affects the stability of the liquid film and subsequently the hot crack formation, resulting in some cases in non-continuous cracks where isolated sections of solid bound the two grains as illustrated by **Figure 3(a)** and **Figure 3(b)**.

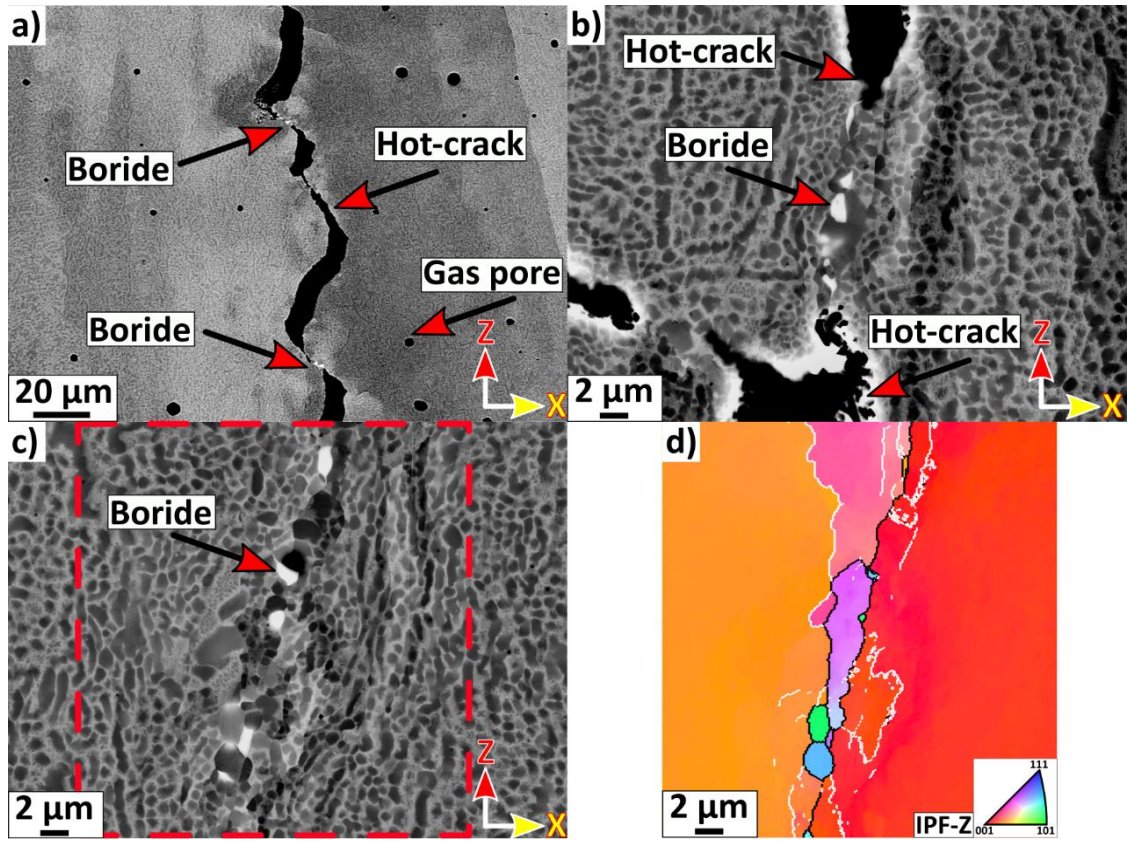

Figure 3: a) Borides forming solid bridges across a high-angle grain boundary in the cracked columnar sample. b) Higher magnification backscattered electron micrograph from a cracked grain boundary of the columnar grain shape sample containing borides. c) Backscattered electron micrograph from a grain boundary of the cracked columnar grain shape sample containing borides and d) EBSD IPF-Z (building direction) map corresponding to the region denoted by the red dashed box in Figure 2c showing recrystallization in the vicinity of intergranular borides.

Table 3: Chemical composition of $M_2B$ boride in the columnar grain shape sample as measured by APT (at.%).

|  | B | Mo | Cr | Ti | Co | Ni |
|---|---|---|---|---|---|---|
| $M_2B$ | 33.86±0.030 | 38.16±0.030 | 21.90±0.026 | 2.11±0.009 | 2.0±0.009 | 1.80±0.008 |

The presence of borides at cracked HAGB is a strong indication that the segregated films contained a higher level of B and boride-formers such as Mo and Cr prior to boride formation. This is confirmed by the APT results in **Figure 4** obtained close to a cracked HAGB (misorientation of 26°) containing several relatively coarse borides. B, C, Mo, and Cr segregate to the grain boundary up to 2.5, 0.5, 10.0 and 30.0 at.%, respectively while Ni, Al and Co are depleted as readily visible in the composition profiles plotted in **Figure 4(b)** and **(c)** respectively. Besides, C and Si were found to segregate to the grain boundary up to 0.5 and 0.2 at.%, respectively, but no specific role on the hot cracking was observed from their presence at the grain boundary. The segregation behavior observed here is in agreement with previous reports [26,32,33]. Segregation of Zr and Hf at the grain boundary was not observed, instead they were found to partition within the γ', as shown in **Supplementary Figure 1(a)**.

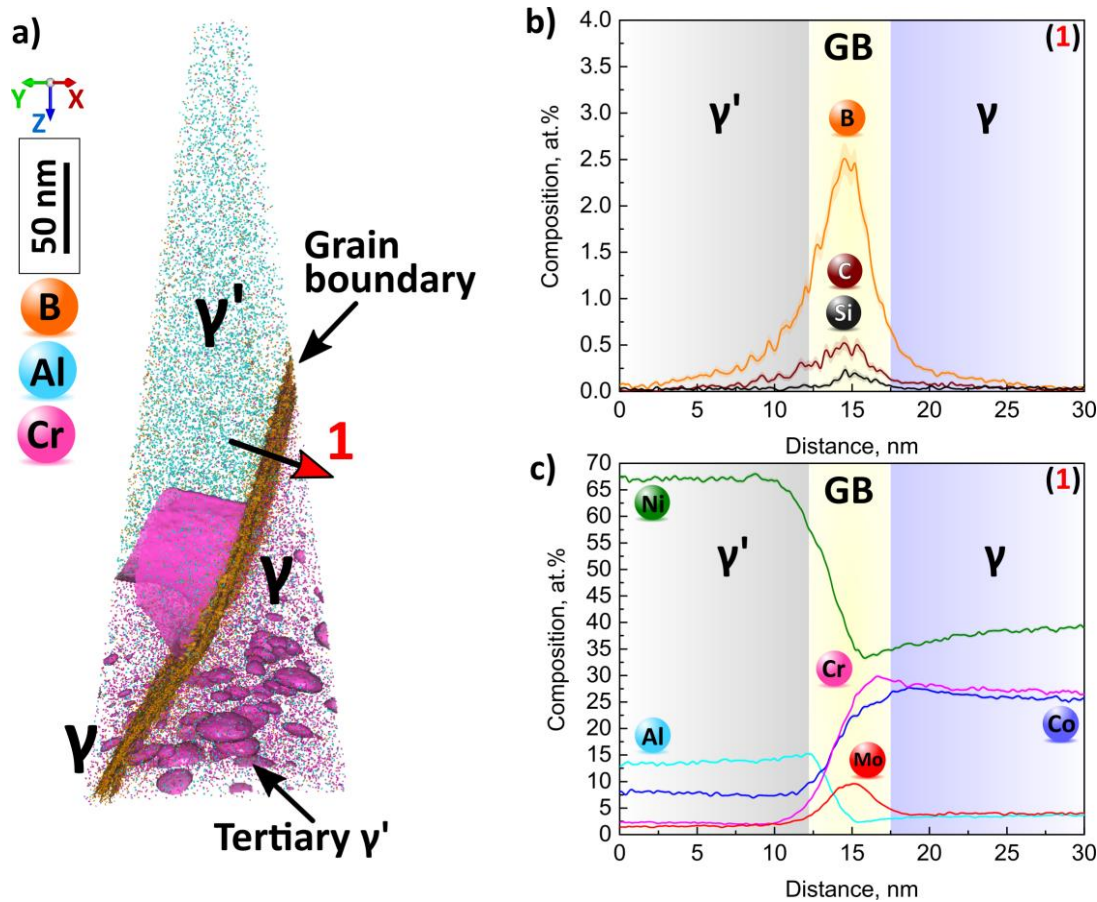

Figure 4: a) Atom probe reconstruction from a random high angle grain boundary in the cracked columnar region. The γ/γ' interfaces are shown by an iso-composition surface encompassing regions in the data containing over 7.0 at.% Cr. The grain boundary is highlighted by an iso-composition surface with a threshold of 1.0 at.%. B and c) One-dimensional composition profiles across the γ'/GB/γ interface as denoted by arrow #1 in Figure 4a are plotted for the minor additions and main alloying elements. Error bars are shown as lines filled with color and correspond to the 2σ counting error.

## 3.2 Microstructural characterization: crack-free sample with equiaxed grains

**Figure 5(a)** shows the sample produced by adjusting the building parameters manually, with full control over the input power, scanning speed and hence melting strategy. The EBSD maps in **Figure 5(b-d)** were obtained approximately at the same height as in **Figure 1**, and they reveal a homogeneous microstructure with equiaxed grains exhibiting a random crystallographic texture with a maximum orientation density of only 1.8 times the random orientation distribution in the {001}-pole figure. **Figure 6(a)** reveals only gas porosity [20] and no lack-of-fusion-like defects, while **Figure 6(b)** is a close-up of the typical γ/γ' microstructure within grains where no borides, no cracks or micro-voids at grain boundaries were found. However, a brightly imaging, discontinuous segregated film appears at HAGBs, shown in **Figure 6(b)**. That film was further investigated by APT.

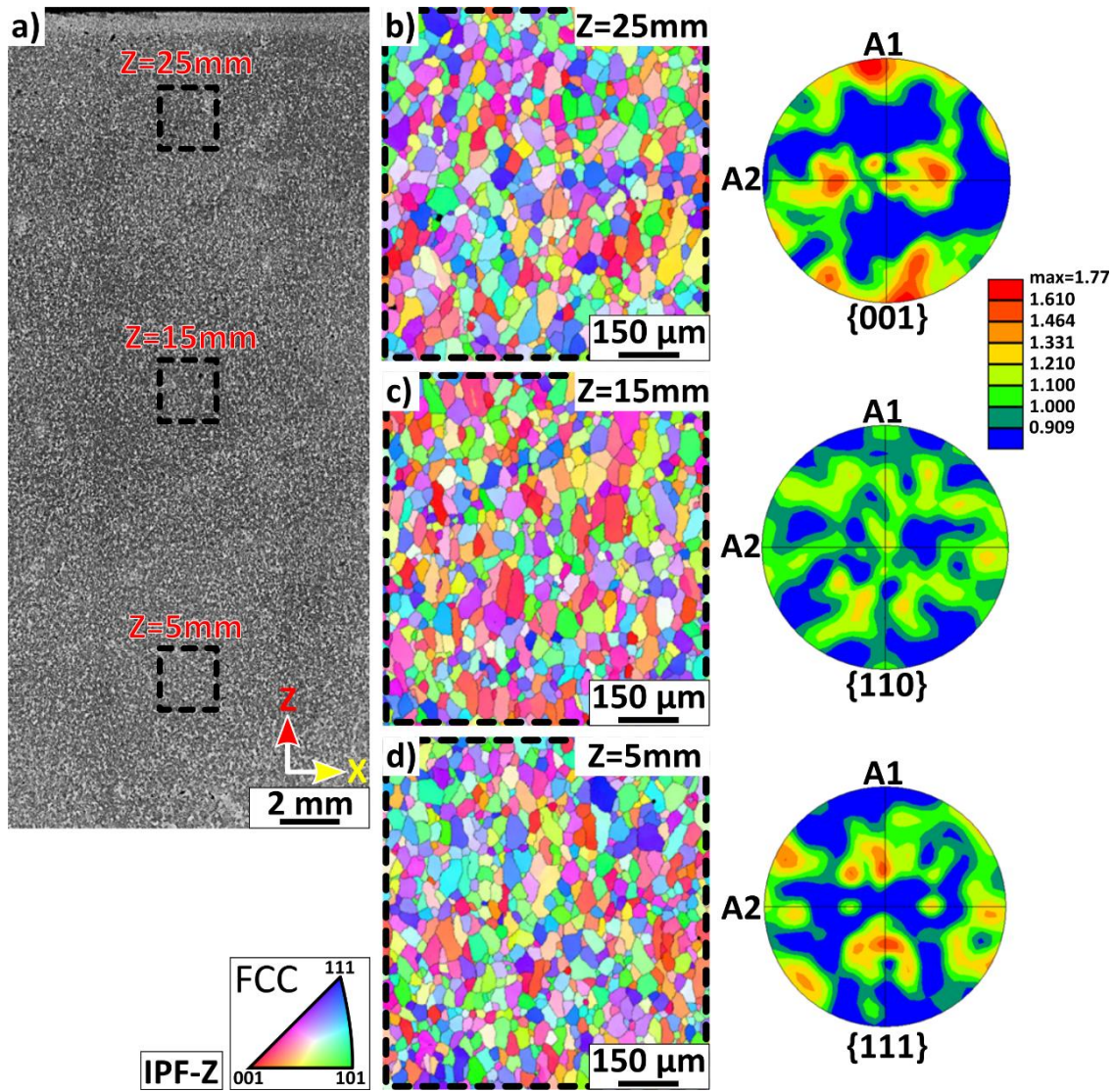

Figure 5: Overview of the crack-free equiaxed sample. a) Optical micrograph of the entire height showing a homogeneous microstructure. b)-d) EBSD IPF-Z (building direction) maps illustrating grains with equiaxed structure. A1 and A2 are the X and Y scanning direction of the electron beam during the deposition.

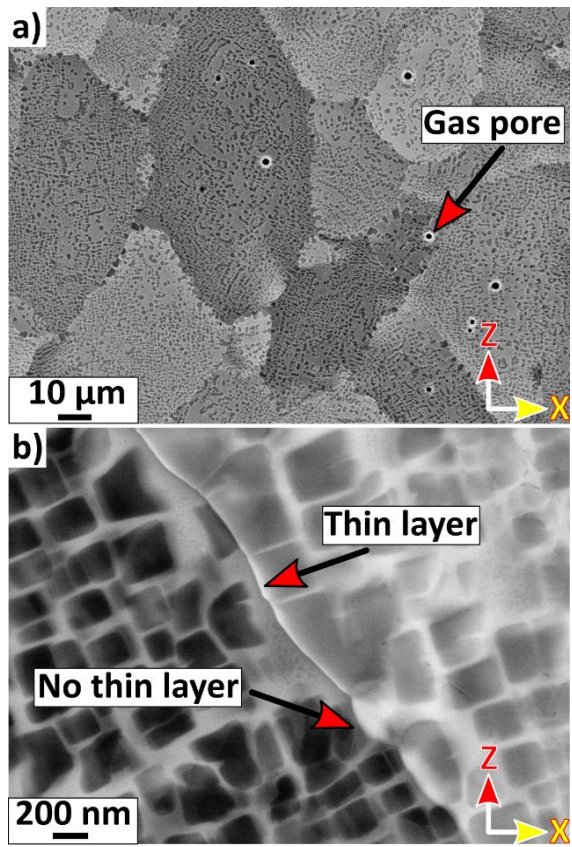

Figure 6: Backscattered electron micrographs from the crack-free with equiaxed grain sample: a) showing only gas pores in the microstructure and absence of intergranular cracks, b) showing a discontinuous segregated film at GBs.

**Figure 7a** shows an APT reconstruction from a HAGB from the crack-free sample with equiaxed grains. TKD was performed on the APT specimen prior to the analysis, shown in **Figure 7(a)**, and confirms the high-angle nature of the grain boundary (53° misorientation). One-dimensional composition profiles across the grain boundary show clear segregation of B, C, Mo, and Cr segregate 7.0, 0.5, 25.0 and 30.0 at.%, respectively. In this case, the amount of B, Mo and Cr at the grain boundary is higher than in the cracked columnar sample (B, Mo, and Cr segregate to the grain boundary up to 2.5, 10.0 and 30.0 at.%) and in samples of a similar grade produced through a powder metallurgical route [29,34].

**Figure 8a** shows a second APT analysis approximately 2 μm farther along the same grain boundary. This is also confirmed by the TKD analysis conducted prior to APT analysis. One-dimensional composition profiles across the grain boundary, as indicated by arrow #1 in **Figure 8a**, reveal segregation of B, C, Mo and Cr up to 4.5, 0.5, 15.0 and 26.0 at.%, respectively. It can be seen that the amount of solutes segregating to the same grain boundary changes over a distance lower than a few micrometers. In particular, in the grain boundary segment shown in **Figure 8**, the presence of the segregated thin film consumes grain boundary solutes resulting in lower amounts of solutes compared to the grain boundary segment being approximately 2 μm farther and that is shown in **Figure 7**. Similar to the columnar cracked sample Zr and Hf were found to partition to γ' and did not segregate at the grain boundary, as shown in **Supplementary Figure 1(b)**.

The segregated film shown in **Figure 8(a)** contains regions with a B content higher than 20.0 at.%, and Cr and Mo at 25–35 at.%, as revealed in **Figure 9(a)**. These are akin to off-stoichiometric boride precursors. **Figures 9(b-d)** are two-dimensional composition maps calculated using the protocol outlined in Ref. [35] for Mo, Cr and B, showing that, away from these precursors, Mo, Cr and B, are only enriched up to 6.0, 11.0 and 2.0 at.% respectively, 3–4 times lower than in **Figure 7(b)** and **Figure 7(c)**. These results evidence how the consumption of solutes by the growth of the borides affects the grain boundary's composition.

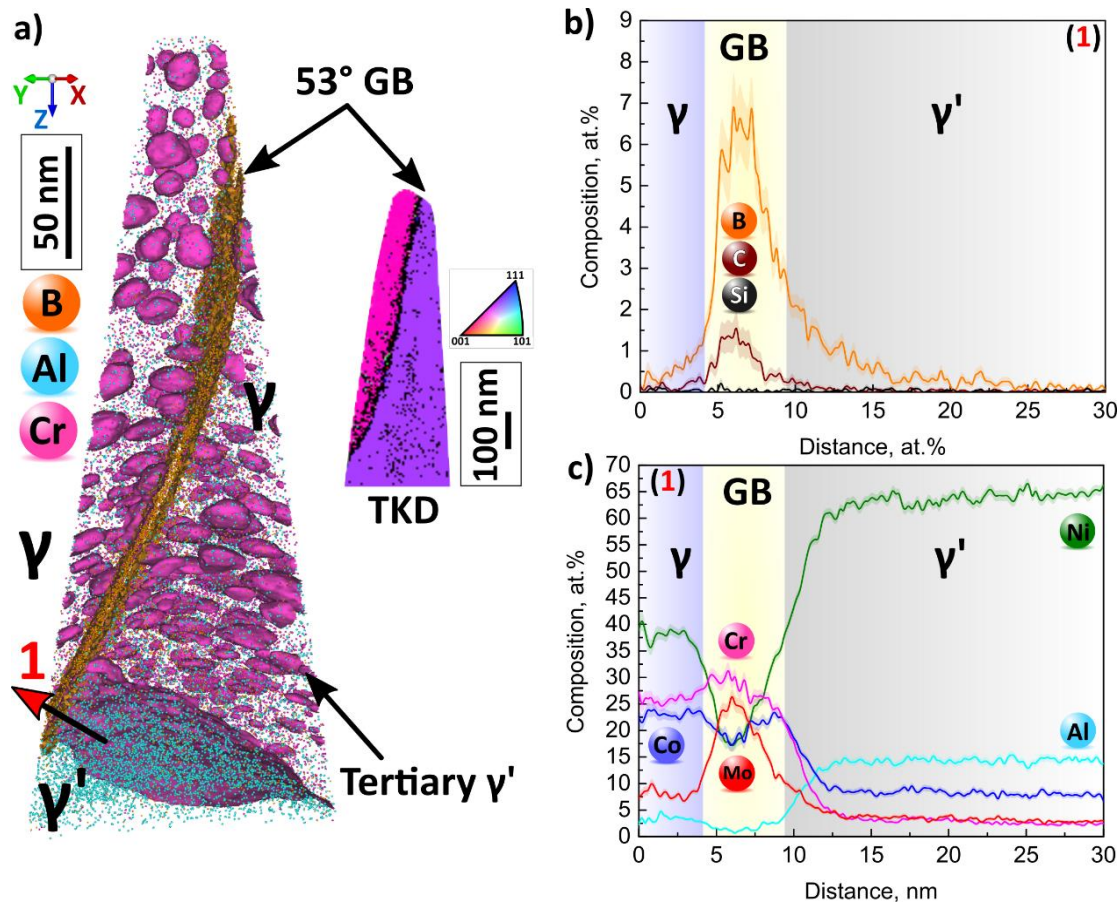

Figure 7: Atom probe reconstruction from a random high angle grain boundary from the crack-free equiaxed grain shape sample alongside an IPF map from TKD. The γ/γ' interfaces are shown with an isosurface with a threshold of 18 at.% Cr and the grain boundary is shown with an isosurface delineating regions containing more than 2.0 at.% B. b)-c) One-dimensional composition profiles across the γ/GB/γ' interface as denoted by arrow #1 in Figure 7a are plotted for the minor additions and main alloying elements. Error bars are shown as lines filled with colour and correspond to the 2σ counting error.

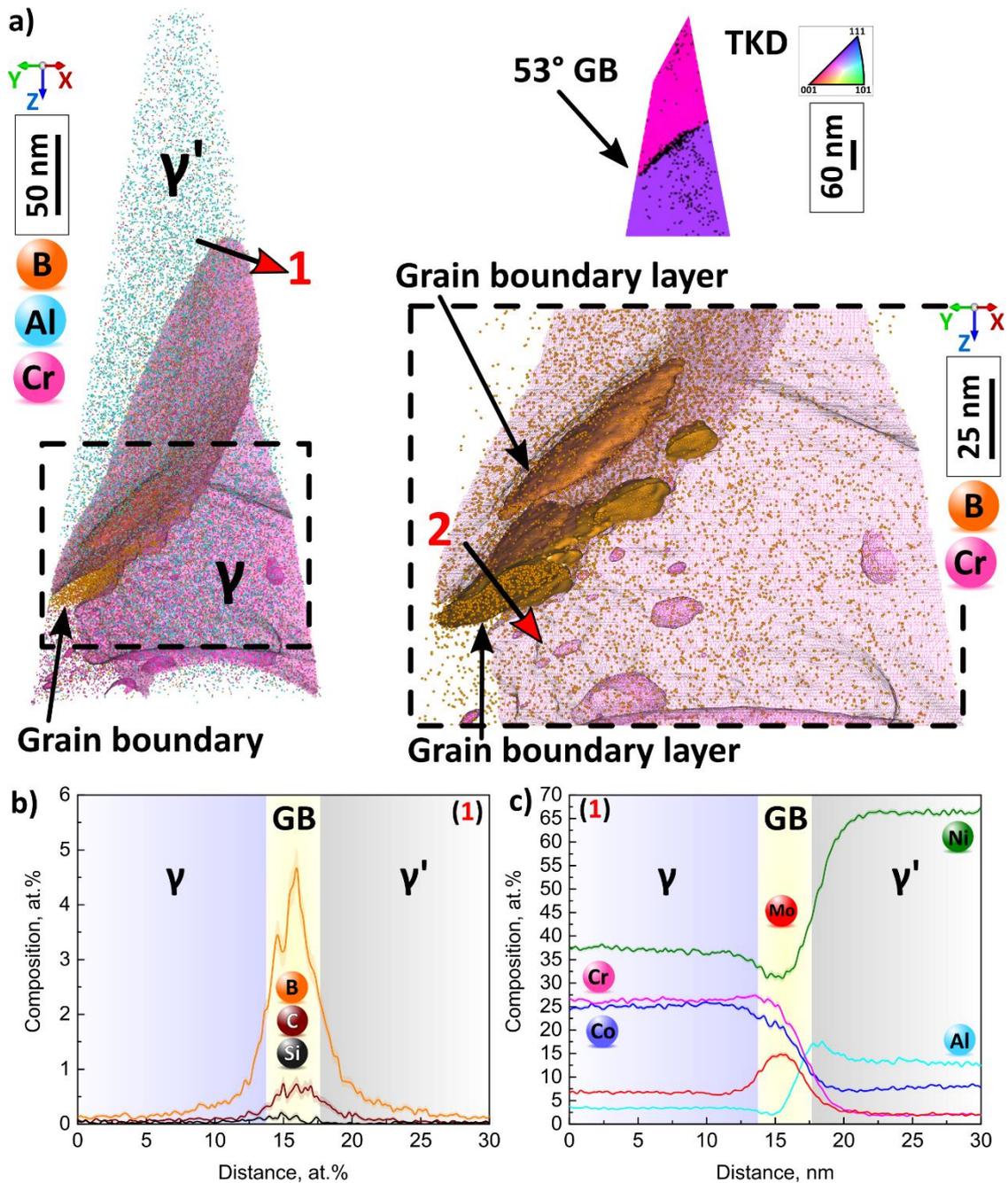

Figure 8: a) Atom probe reconstruction from the cracked-free equiaxed grain shape sample and the same random high angle grain boundary as in Figure 7 and IPF map from the TKD analysis. The γ/γ' interfaces are shown with an isosurface with a threshold of 7 at.% Cr and the segregated film is shown with an isosurface at 10.0 at.% B. b)-c) One-dimensional composition profiles across the γ/GB/γ' interface as denoted by arrow #1 in Figure 8a are plotted for the minor additions and main alloying elements. Error bars are shown as lines filled with colour and correspond to the 2σ counting error.

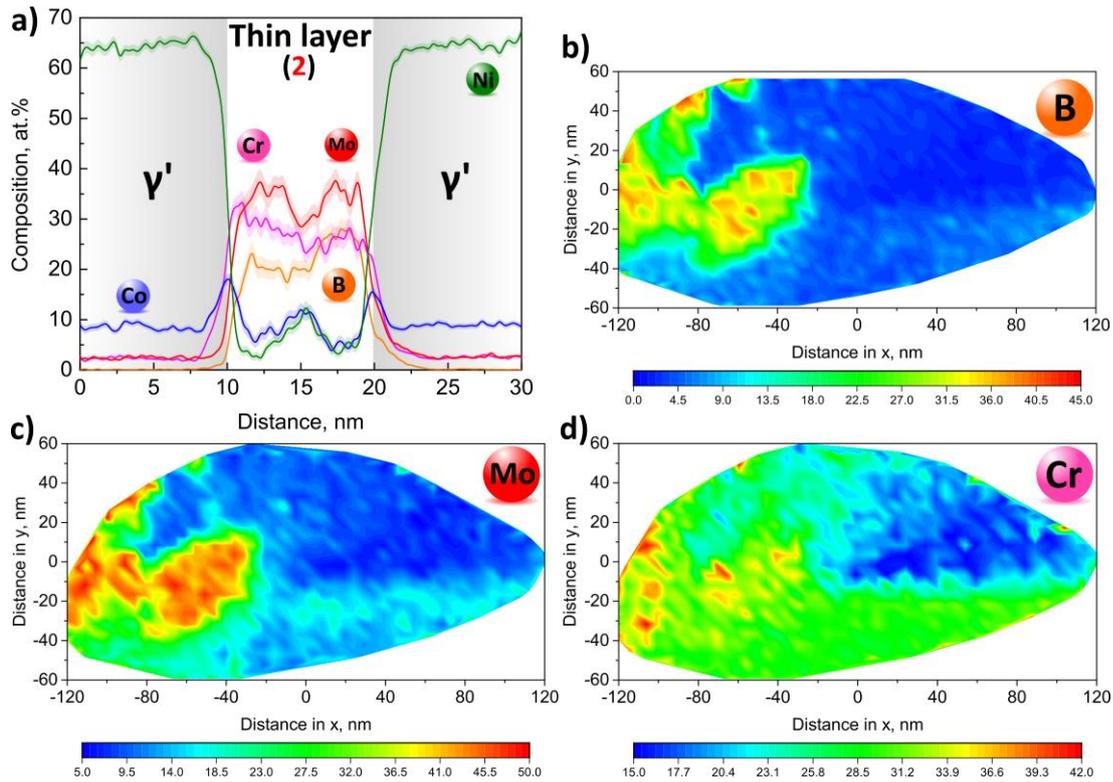

Figure 9: a) 1D composition profiles across the GB along arrow #2 across the boride precursor in Figure 8a. Error bars are shown as lines filled with colour and correspond to the 2σ counting error. b)-d) 2D compositional maps calculated inside the grain boundary plane for Mo, Cr and B. Note that the color-bar scale gives the solute content in at.%.

However, these thin discontinuous grain boundaries segregated films are not as detrimental as the hot cracks observed in the cracked columnar sample, as revealed by small-scale tensile tests performed on specimens machined from the as-produced samples. **Figure 10** shows the engineering stress vs. engineering strain curves resulting from 3 tensile tests for the cracked columnar and crack-free equiaxed microstructures. The coarse-grained columnar samples show a very brittle behavior, indicative of samples that already contain critical defects. The tensile tests performed on the crack-free samples with equiaxed grain structure show a vastly improved behavior from approximately 5.0 % in the cracked columnar sample up to between 20.0 and 27.0 % engineering strain, with no tendency for early fracture, due to the absence of hot cracks.

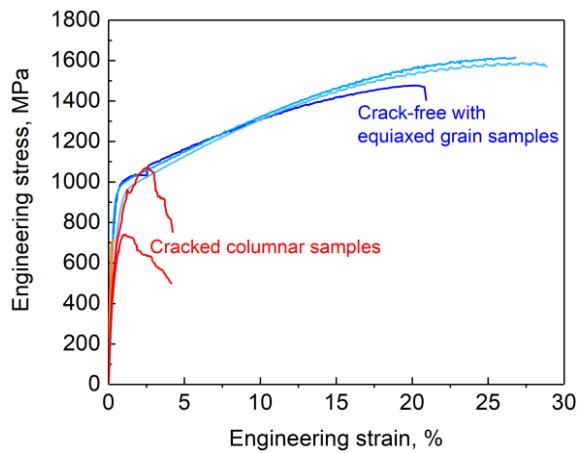

Figure 10: Engineering strain vs. engineering stress curves. Three micro-specimens with a gauge length of 3 mm, a width of 1 mm, and a thickness of 0.8 mm were tested for the cracked columnar microstructure and three others for the crack-free equiaxed microstructure. The load axis was normal to the building direction Z.

### 3.3 Thermodynamic calculations

To evaluate the influence of the content of B, Cr and Mo on the solidification path, thermodynamics calculations were performed using Thermo-Calc software making use of the TTNI8 database dedicated to nickel-based superalloys [28]. Calculations of the solidification path based on the Scheil-Gulliver model were performed including the liquid, γ and γ' phases and excluding or including the formation of $M_2B$ borides, as reported in **Figure 11(a)** and **(b)**, respectively. No diffusion in the solid state and infinite diffusion in the liquid were assumed in the Scheil-calculations. The calculations allow to simulate sequential steps from the liquidus temperature to an approximate solidus temperature, giving the fraction of solid and composition of the new liquid at each temperature. The comparison between **Figure 11(a)** and **(b)** shows that liquid films remain stables at lower temperatures when considering that borides are not a solidification product but rather form in the solid state. The solidus is 60°C lower when excluding borides from the calculations. The composition of the liquid as a function of the temperature of the investigated alloy is plotted in **Figure 11(c)** as a function of the mole fraction of B, Cr and Mo respectively in orange, pink and red when assuming that borides are not a solidification product (boride not considered in the calculation). The three solutes, namely B, Cr and Mo strongly partition to the liquid, and the liquid is hence enriched in those solutes as the solidification proceeds, reaching approximately 6.0, 19.0 and 24. at.% respectively during the last stage of solidification.

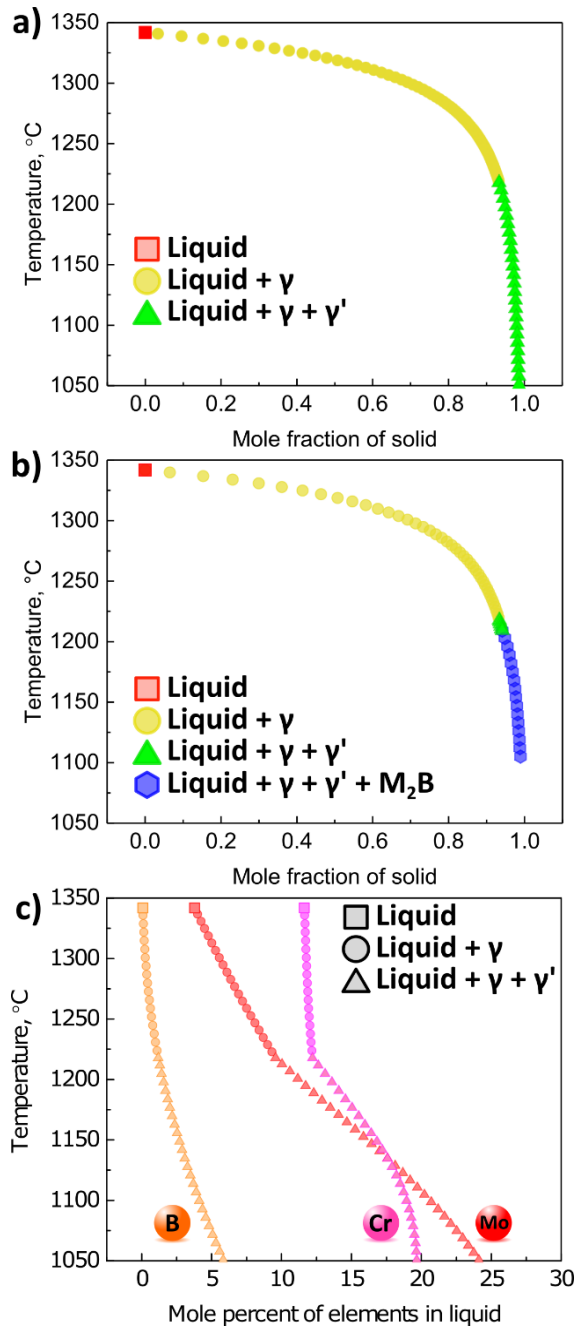

Figure 11: Solidification paths relying on the Scheil-Gulliver assumptions (a) without and (b) with the precipitation of $M_2B$ borides as a solidification product. Calculations performed using Thermo-Calc and relying on TTNI8 database. c) Evolution of the B-, Cr- and Mo-content as a function of the temperature in the liquid phase when considering only γ, γ′ and liquid (excluding borides).

# 4 Discussion

## 4.1 Grain boundary segregation and boride formation

Segregation to grain boundaries may originate in AM first from solidification by sweeping of solutes during grain growth as new layers are deposited, and second by solid-state diffusion as the sample is maintained at high temperature (in S-EBM a heated powder bed is used). The intergranular films get enriched in solutes, such as Cr, Mo and B, during the building process, because of remelting and most likely solid-state diffusion, particularly of fast diffusers such as

B [26]. HAGBs accommodate larger amounts of solutes than LAGBs [35], partly due to lower their relatively higher interfacial energy [36].

**Figure 4**, **Figure 7** and **Figure 8** show clear segregation of B, Mo, and Cr to the grain boundaries. Under non-equilibrium conditions, such as experienced during S-EBM, thermodynamic calculations of the first solidification event predict a substantial enrichment of B, Mo and Cr in the liquid as the temperature decreases, as reported in **Figure 11(c)**. The enrichment predicted by the Scheil-Gulliver calculations is however likely insufficient to form borides during the first solidification event, which is supported by the absence of borides in the last deposited layer. Upon several remelting events, as B gets more enriched and thus concentrated, a critical composition of boride-formers is finally reached and borides can nucleate. They can likely form in the liquid or heterogeneously in the solid state, and coarsen quickly as HAGBs are fast diffusing paths.

The small off-stoichiometric borides observed in the sample with equiaxed grains are an indication that the borides form and grow progressively during the build, and those observed here, albeit small, resulted from multiple reheating/remelting events. These small borides can thus be interpreted as early-stage variants of the larger borides observed in the coarse-grained columnar sample. The composition profile shows an enrichment of Co up to 18.0 at.% at the interface between γ' and the boride-precursor compared to 10.0 at.% within the γ' precipitate and 5.0-10.0 at.% within the off-stoichiometric borides. This indicates outward Co diffusion from the boride towards the γ' precipitate. Fully developed borides analyzed in the coarse-grained columnar sample contain only approx. 2 at.% Co, while these precursors still contain approximately 5.0 to 10.0 at.%. Also these findings support the assumption that we observed here a transient state. In cases where borides do not deplete the film in B, Mo and Cr, the high content in these boride-forming solutes in the last solidified liquid films lowers the alloy's solidus potentially significantly, as indicated by the results reported in **Figure 11(a)** and **11(c)**.

The growth of borides consumes solutes, which, as indicated by thermodynamic calculations, increases the local solidus near the boride by locally decreasing the content of solutes such as B and Mo, as proven in **Figure 9**. In such a case, isolated sections of solid binding between two adjacent grains appear within the liquid film, as revealed in **Figure 3(a)**. This finding illustrates the direct influence of the local partition effects and hence of the local solute composition content on the solidification, and hence on the cracking susceptibility.

### 4.2 Hot cracking caused by grain boundary-segregation induced liquation

**Figure 1(f)** gives strong indication that cracks are caused by interfacial liquation along some of the grain boundaries and **Figure 2** demonstrates that the cracks do not originate directly from the first solidification event since no cracks propagate through the last melted layer. Thermodynamic calculations demonstrate a progressive enrichment in critical solutes within the liquid film during the last stage of the first solidification event. The characterization by APT of the films observed along HAGBs confirms a significant enrichment in these same critical solutes, namely Cr, Mo and B, which can be further accentuated by both successive remelting and solid-state diffusion. This is particularly true for fast diffusing elements, such as B.

Regarding the cracking mechanism evolving during the building process, over the course of the deposition of the subsequent layers, a segregated film at a HAGB can turn into a liquid film. The drop in the solidus caused by the progressive enrichment in solute is referred to as

"segregation induced liquation" [37]. The difference in volume between the liquid and solid phase associated with the solidification, i.e. the shrinkage effect, as well as thermal stresses caused by rapid cooling provide the mechanical driving forces for the development of hot cracks [38]. These local thermal stresses combined with segregation-driven grain boundary liquation cause hot cracking [39,40]. The building history of the cracked columnar sample that leads to cracking is summarised in the schematic in **Figure 12**. Ultimately, hot cracking requires a condition setting where concentration in the liquid reaches a critical value, which appears close to the critical composition to form borides, in combination with the solidification stresses.

B is usually added to superalloys to increase GB cohesion [41] and limit damage at GBs under creep conditions. However, here it acts detrimental and promotes hot cracking as it depresses the solidus temperature, hence stabilize liquid films to lower temperatures. The limited number of grain boundaries in columnar samples will tend to have a more highly concentrated and continuous films extending over several grains, which appears to be highly detrimental to the structural integrity of the build. The typical criterion for non-weldability, namely the content of (Ti+Al) [2], does not seem to apply here. In the alloy that we investigated, this criterion is superseded by the content of B, Cr and Mo found at HAGBs. Ultimately, HAGBs crack because (i) the films are continuous and extend over several mm and (ii) the local composition of the segregated films allows the liquid to remain stable to relatively lower temperatures, hence causing grain boundary liquation during the deposition of the subsequent layers. Our observations imply that a bulk-criterion may not allow to appropriately depict near-atomic-scale processes.

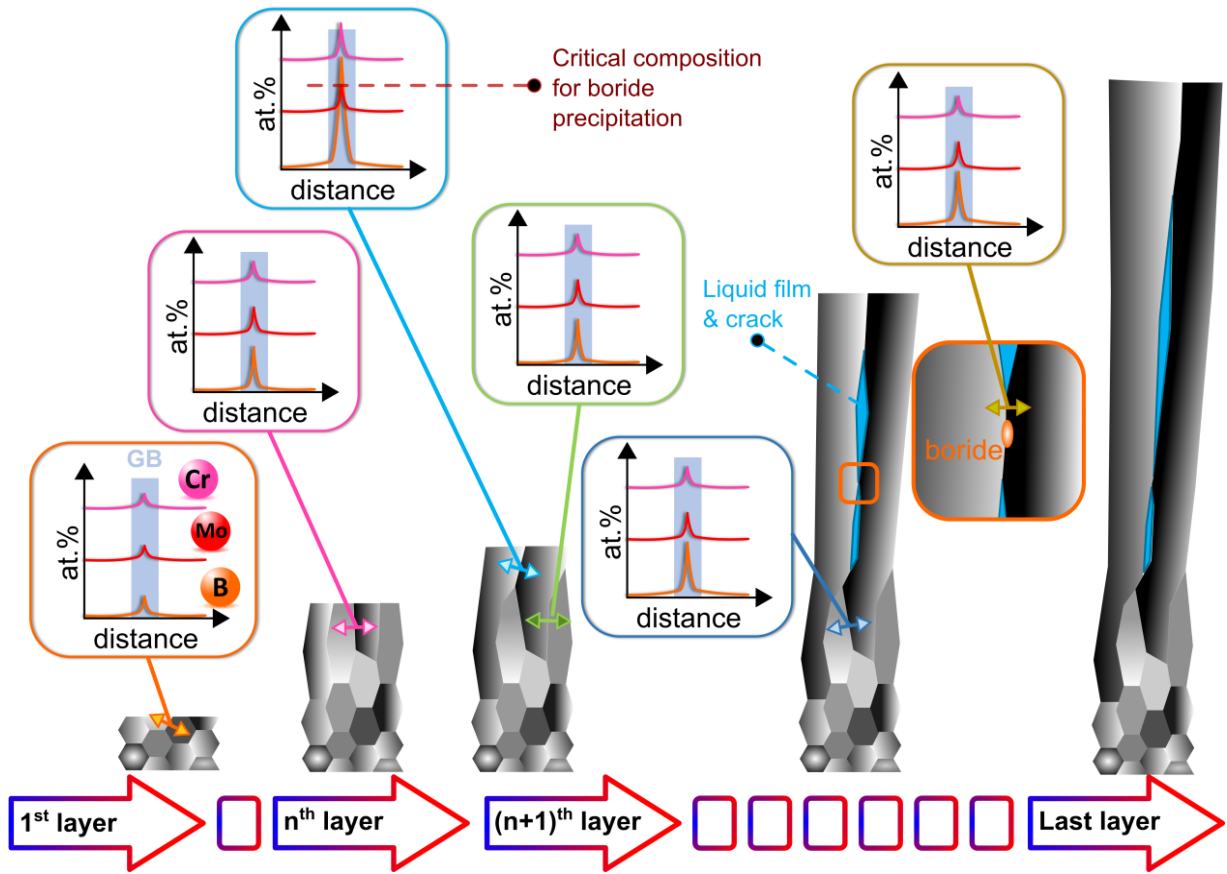

Figure 12: Schematic summary of the microstructural and compositional evolution during the build of the sample with hot cracks.

It is also known that Zr promotes hot cracking in nickel-based superalloys produced by casting [42–45] and AM processes [5]. Although Zr is added in the bulk composition of the investigated alloy (0.02 – 0.025 at.%), APT revealed no segregation of Zr at the grain boundary. Our observation on Zr is consistent with other studies performed by Cadel et al. [29] and Lemarchand et al. [34]. By contrast, Zr was found to partition to the γ' phase for both columnar cracked and equiax crack-free samples, as shown in **Supplementary Figure 1**. However, in a different nickel-based superalloy, Zr was found to segregate at the grain boundary [5]. Besides, significant interactions between Zr and B were recently reported during casting [44]. In order to rationalize the interactions between the different alloying elements in such chemically complex metallic systems, further detailed investigations are required which are outside the scope of this present work.

### 4.3 Strategies for additive manufacturing of crack-free superalloys components

Better understanding and quantitative mapping of solute segregation to grain boundaries allows us to derive strategies to avoid cracking in additively manufactured superalloy parts. Modulating solute segregation at grain boundaries allows to better control liquation and better distribute the thermal stresses associated to solidification and rapid cooling. This can be achieved by controlling the total interfacial area. This means that, depending on the bulk concentration of critical solutes, one has to adjust the overall interfacial area to gain control over their segregation to HAGBs in order to avoid liquation.

Interestingly, our APT results show higher solute enrichment along HAGBs in the crack-free sample with equiaxed grains compared to the ones in the cracked columnar sample. At first sight, it could be considered as not consistent with our argument stating that increasing the interfacial area should result in a less concentrated thin segregated films in critical solutes, i.e. the ones reducing the solidus temperature. The distribution of solutes at larger interfacial are would then allow to limit critical grain boundary segregation that can lead to liquation in presence of thermal stresses. However, those results have to be considered in light of the presence of coarse micron-sized borides that consume large quantities of critical solutes such as B, Mo and Cr. In the cracked sample with columnar grains we have extracted APT specimens in the immediate vicinity of coarse borides. We show how the precipitation of borides along HAGB and the associated partitioning of grain boundary solutes into these particles can affect locally the grain boundary segregation. Our results suggest that prior to boride formation, HAGBs were likely decorated by a continuous and highly segregated film, which increases the liquation susceptibility. However, this contradicts our observations at HAGBs in the sample with equiaxed grains, where no borides and no cracks were observed, only discontinuous segregated films. In the equiaxed sample, the grain boundary enrichment in B, Cr and Mo had also slowed down as solutes are spread over a wider area and must diffuse over longer distances to reach a high local composition. Conversely to the cracked columnar sample, the grain boundary area in the equiaxed samples translates into discontinuous segregated films as evidenced in **Figure 6(b)**. In addition, the presence of more interfaces in the equiaxed version helps to better accommodate solidification shrinkage and thermal stresses, hence the likelihood of crack initiation is significantly reduced in comparison with a microstructure made of coarse columnar grains.

It becomes apparent that a strategy for AM produced crack-free superalloys should have good control of the total interfacial area of the produced component. This will prevent from reaching the critical composition for the formation of detrimental continuous liquid films and undesirable local thermal stresses over very few interfaces.

Based on this strategy aiming at controlling carefully the interfacial area, it would be possible to produce a crack-free columnar sample, by once again taking advantage of the digital control of the melting strategy offered by S-EBM, to achieve a critical grain width that leads to crack-free samples. **Figure 13** shows the evolution of grain width as a function of building height for the cracked columnar and crack-free equiaxed sample. It can be seen that for a grain width of approximately 20 μm, the equiaxed sample is crack free. Interestingly, the cracked columnar is also free of cracks over the first roughly 10 millimetres of the build which corresponds to a grain width lower than 100 microns. In other words, by producing a finer columnar microstructure with grain width < 100 μm, one should get samples free of cracks.

We have produced a third sample which was built by using a relatively low power and slow scan speed (linear energy of 0.3 J/mm²) while maintaining a line offset of 0.1 mm. The low power maintains a relatively small melt pool that hinders grain coarsening and limits grain selection. As a result, the grain width remains small, approximately 45 μm as shown in **Figure 13** along the entire height of the build. An overview of this fine columnar microstructure shows no hot cracks, as revealed in **Figure 14**.

A possible alternative route would consist in adjusting the composition to achieve lower level of solutes along HAGBs that decrease the solidus for instance or to form a continuous layer of e.g. borides to strengthen the grain boundaries [26,46].

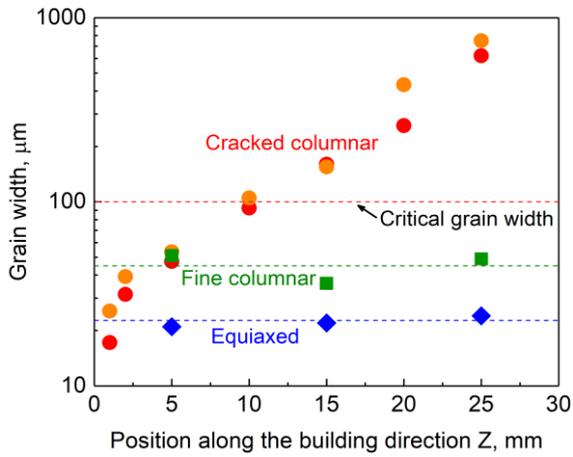

Figure 13: Quantification of grain width along the building direction for cracked and crack-free samples. Results from two different cracked columnar samples were included to highlight that the grain width is reproducible from one build to another. Blue and green dashed lines represent the average grain width for the sample with equiaxed and columnar grains, respectively. Red dashed line corresponds to the critical grain width, beyond that hot cracks occur.

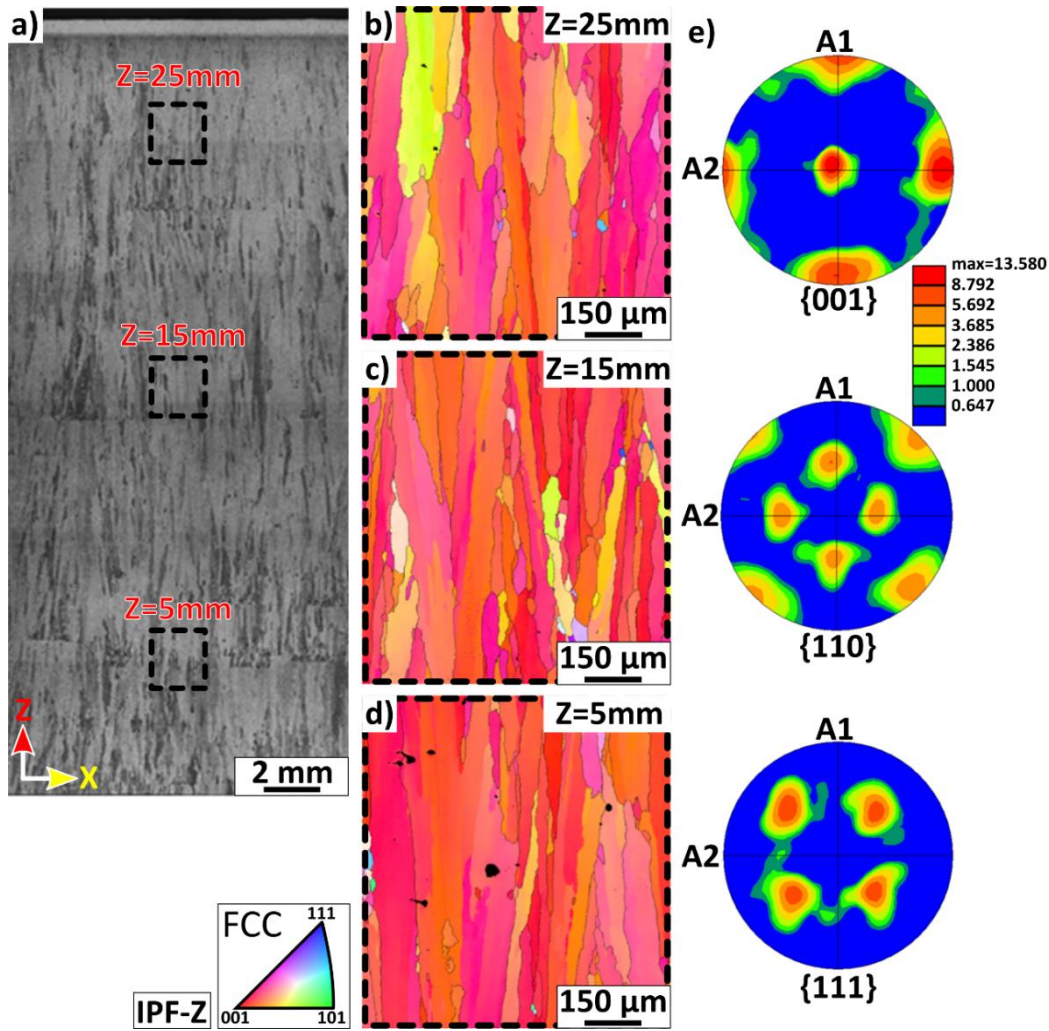

Figure 14: Overview of the crack-free fine-grained columnar sample: optical micrograph along with detailed EBSD IPF-Z (building direction) maps and pole figure illustrating the fine columnar grains along the building direction. A1 and A2 are the X and Y scanning direction of the electron beam during the deposition.

## 5. Conclusions

To conclude, we combined microstructural observations by optical and electron microscopy with near-atomic scale compositional investigation of critical microstructural features by atom probe tomography, and thermodynamic calculations to study additively manufactured builds from a non-weldable nickel-based superalloy produced by S-EBM. Our approach allowed us to provide fundamental insights into the dynamic interplay, at the nano-scale, of the local composition in critical regions, such as HAGBs, that leads to segregation-induced liquation cracking.

The new fundamental understanding of the root cause of cracking set us on the path to tune the atomic-scale distribution of critical solutes, such as B, Cr and Mo, in particular along HAGBs. This was made possible by the high level of control over the melting strategy offered by the S-EBM process. Through modifications in the melting strategy, we adjusted the geometry of the melt pool which turned the coarse-grained columnar microstructure sensitive to liquation cracking into either small grained equiaxed (grain width approximately 20 μm) or fine columnar microstructures (grain width approximately 50 μm). Increasing the interfacial area not only prevents high concentration of boride-forming elements that cause liquation, and hence avoids cracking but also helps to distribute thermal stresses over more interfaces.

**Acknowledgements**

This work was performed within the framework of the Center of Excellence of Multifunctional Architectured Materials ''CEMAM'' n°AN-10-LABX-44-01 funded by the ''Investments for the Future Program''. The authors are grateful to Poly-Shape for their contribution to funding this work. We thank Uwe Tezins & Andreas Sturm for their support to the FIB & APT facilities at MPIE. We are grateful for the financial support from the Max-Planck Gesellschaft via the Laplace project for both equipment and personel (PK, AKdS). ZP acknowledges financial support from the BIGMAX project. JH is grateful for funding by the DFG through the SFB TR103. BG acknowledge financial support from the ERC-CoG-SHINE-771602.

**Author contributions**

P.K., B.G., J-J.B and G.M. designed the study. E.C. and G.M. 3D-printed the alloy. P.K., B.G. and D.R. prepared the atom probe specimens, processed the data and interpreted the results. Z.P. conducted the grain boundary analysis. J.H. and P.K. conducted the TKD analysis. A.K.S. performed Thermo-Calc calculations, C.T. provided support on these. B.G., P.K. and G.M. wrote the main draft of the paper. All authors contributed to the discussion of the results and commented on the manuscript.

of boron in γ/γ′ Cobalt-base superalloys, Acta Mater. 145 (2018). doi:10.1016/j.actamat.2017.12.020.

**Supplementary Figure**

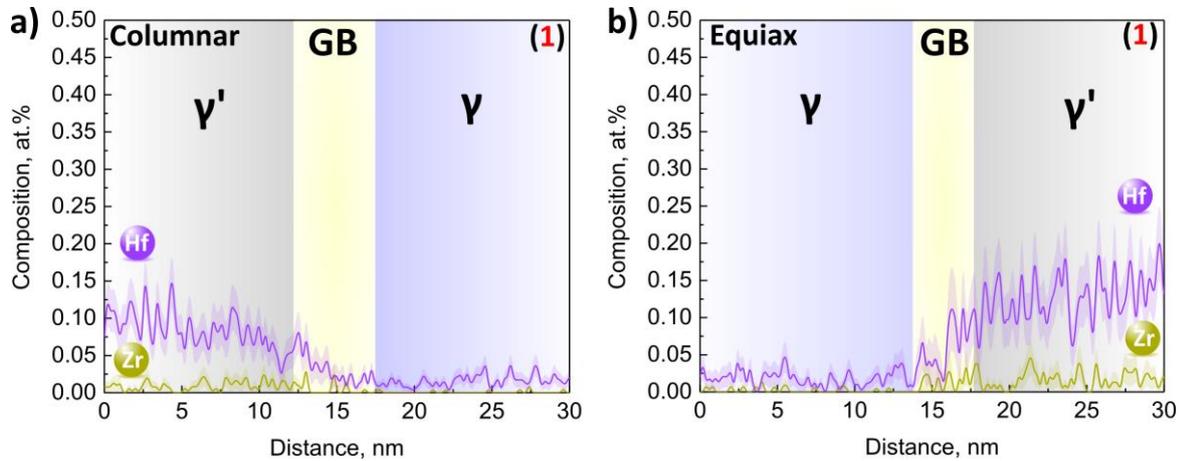

Supplementary Figure 1: a) One-dimensional composition profiles across the γ'/GB/γ interface as denoted by arrow #1 in Figure 4a from the cracked columnar sample are plotted for Zr and Hf. b) One-dimensional composition profiles across the γ/GB/γ' interface as denoted by arrow #1 in Figure 8a from the crack-free equiaxed sample are plotted for Zr and Hf. Error bars are shown as lines filled with colour and correspond to the 2σ counting error.